\documentclass[twocolumn]{emulateapj} 

\usepackage{amsfonts}
\usepackage{amsbsy}
\usepackage{natbib}
\usepackage{graphicx}
\usepackage{ulem}

\usepackage{color}

\newcommand{\myvec}[1]{\boldsymbol{#1}}
\newcommand{\mymat}[1]{\boldsymbol{#1}}
\newcommand{\Map}{\ensuremath{M_{\rm ap}}\ }
\newcommand{\naive}{na\"{i}ve\ }
\newcommand{\Noise}{\mymat{\mathcal{N}}}

\slugcomment{\textit{Draft Version, \today.} }
 
\alph{footnote}

\begin{document}

\title{Interpolating Masked Weak Lensing Signal
  with Karhunen-Lo\`{e}ve Analysis}

\author{J. T. VanderPlas and A. J. Connolly}
\affil{Astronomy Department, University of Washington, Box 351580, 
  Seattle, WA 98195-1580}
\and
\author{B. Jain and M. Jarvis}
\affil{Department of Physics and Astronomy, University of Pennsylvania, 
  209 South 33rd Street, Philadelphia, PA 19104-6396}
\bibliographystyle{apj}

\begin{abstract}
  We explore the utility of Karhunen Lo\`{e}ve (KL) analysis in 
  solving practical problems in the analysis of gravitational
  shear surveys.  Shear catalogs from large-field weak lensing
  surveys will be subject to many systematic limitations, notably
  incomplete coverage and pixel-level masking due to foreground sources.  
  We develop a method to use two dimensional KL eigenmodes of 
  shear to interpolate noisy shear measurements across masked regions.  
  We explore the results of this method with simulated shear catalogs, 
  using statistics of high-convergence regions in the resulting map.  
  We find that the KL procedure not only
  minimizes the bias due to masked regions in the field, it also reduces
  spurious peak counts from shape noise by a factor of $\sim 3$ in the
  cosmologically sensitive regime.  This indicates that KL reconstructions 
  of masked shear are not only useful for creating robust convergence maps
  from masked shear catalogs, but also offer promise of improved parameter
  constraints within studies of shear peak statistics.
\end{abstract}


\section{Introduction}
Since its discovery over a decade ago, cosmic shear -- the coherent 
gravitational distortion of light from distant galaxies due to large
scale structure -- has become an important tool in precision cosmology.
\citep[see][for a review]{Bartelmann01}.  
\citet{Kaiser95} first proposed a method of using weak gravitational lensing
to probe the projected gravitational mass within an observed field.  This 
method and its variations have often been applied to reconstructions of
clusters using deep observations within small (a few square arcminute) fields.
In this situation, the measured shear often dominates the shape noise, and 
useful priors can be applied based on assumptions regarding 
the location of the cluster center, symmetries of cluster mass profile, 
and the correlation of mass with light.

Currently, a new generation of wide-field weak lensing surveys are
in the planning and construction stages.  
Among them are the the Dark Energy Survey (DES), 
the Panoramic Survey Telescope \& Rapid
Response System (PanSTARRS), the Wide Field Infrared Survey Telescope (WFIRST),
and the Large Synoptic Survey Telescope (LSST), to name a few.
These surveys, though not as deep as small-field space-based lensing surveys,
will cover orders-of-magnitude more area on the sky: up to $\sim 20,000$
square degrees in the case of LSST.

This is a fundamentally different regime 
than early weak lensing reconstructions
of single massive clusters: the strength of the shear signal is only
$\sim 1$\%, and is dominated by $\sim 30$\% intrinsic shape noise.  
This, combined with source galaxy densities of only 
$n \sim 20-50\ \mathrm{arcmin}^{-2}$
(compared with $n > 100\ \mathrm{arcmin}^{-2}$ for deep, space-based surveys)
and atmospheric psf effects leads to a situation where the signal is 
very small compared to the noise.  
Additionally, in the wide-field regime,
the above-mentioned priors cannot be used.  
Nevertheless, many methods have been developed to extract useful
information from wide-field cosmic shear surveys, including
measuring the N-point power spectra and correlation functions 
\citep{Schneider02,Takada04,Hikage10}, 
performing log transforms of the convergence field 
\citep{Neyrinck09,Neyrinck10,Scherrer10,Seo11},
analyzing statistics of convergence and aperture mass peaks 
\citep{Marian10,Dietrich10,Schmidt10,Kratochvil10,Maturi11}.
Another well-motivated application of wide-field weak lensing 
is using wide-field mass reconstructions to minimize the effect
mass-sheet degeneracy in halo mass determination.

Many of the above applications require reliable recovery of 
the projected density, 
either in the form of the convergence $\kappa$, or filter-based quantities 
such as aperture mass \citep{Schneider98}.  
Because each of these amounts to a non-local filtering of the shear, 
the presence of masked regions can lead to a bias across significant
portions of the resulting maps.  Many of these methods have been demonstrated 
only within the context of idealized surveys, with exploration of the 
complications of real-world survey geometry left for future study.  
Correction for masked pixels has been studied within the context 
of shear power spectra \citep{Schneider10,Hikage10}
but has not yet been systematically addressed
within the context of mapmaking and the associated statistical methods
\citep[see, however,][for some possible approaches]{Padmanabhan03,Pires09}.
We propose to address this missing data problem through 
Karhunen-Lo\`{e}ve (KL) analysis.

In Section~\ref{KL_Intro} we summarize the theory of KL analysis in the
context of shear measurements, including the use of KL for interpolation
across masked regions of the observed field.
In Section~\ref{Testing_Reconstruction} we show the shear eigenmodes for
a particular choice of survey geometry, and use these eigenmodes to
interpolate across an artificially masked region in a simulated shear catalog.
In Section~\ref{Shear_Peaks} we discuss the nascent field of 
``shear peak statistics'',
the study of the properties of projected density peaks, and propose this
as a test of the possible bias imposed by KL analysis of shear.
In Section~\ref{Discussion} we utilize simulated shear catalogs 
in order to test the effect of KL interpolation on
the statistics of shear peaks.

\section{Karhunen-Lo\`{e}ve Analysis of Shear}
\label{KL_Intro}
KL analysis is a commonly used statistical tool
in a broad range of astronomical applications, from, e.g.~studies of 
galaxy and quasar spectra \citep{Connolly95,Connolly99,Yip04A,Yip04B}, to 
analysis of the spatial distribution of galaxies 
\citep{Vogeley96,Matsubara00,Pope04}, to characterization of the 
expected errors in weak lensing surveys \citep{Kilbinger06, Munshi06}.    
Any set of $N$-dimensional data can be represented as a sum of 
$N$ orthogonal basis functions: this amounts to a rotation and scaling of 
the $N$-dimensional coordinate axis spanning the space in which the data live.
KL analysis seeks a set of orthonormal basis functions which can optimally
represent the dataset.  The sense in which the KL basis is optimal will be
discussed below.  For the current work, the data we wish to represent are the 
observed gravitational shear measurements across the sky.  
We will divide the survey 
area into $N$ discrete cells, at locations $\myvec{x}_i,\ 1\le i \le N$.  
From the ellipticity of the galaxies within each cell, 
we infer the observed shear $\gamma^o(\myvec{x}_i)$, which we assume
to be a linear combination of the true underlying shear, $\gamma(\myvec{x}_i)$
and the shape noise $n_\gamma(\myvec{x}_i)$.\footnote{
Throughout this work, we assume we are in the regime where the convergence
$\kappa \ll 1$ so that the average observed ellipticity in a 
cell is an unbiased estimator of shear; see \citet{Bartelmann01}}
In general, the cells may be of any shape (even overlapping) 
and may also take into account the redshift of sources.
In this analysis, the cells will be square pixels across the locally 
flat shear field, with no use of source redshift information.  
For notational clarity, we will represent quantities with a vector notation,
denoted by bold face: i.e. $\myvec{\gamma} = [\gamma_1,\gamma_2\cdots]^T$; 
$\gamma_i = \gamma(\myvec{x}_i)$. 

\subsection{KL Formalism}
\label{KL_Formalism}
KL analysis provides a framework such that our measurements $\myvec\gamma$ 
can be expanded in a set of $N$ orthonormal basis functions 
$\left\{ \myvec{\Psi}_j(\myvec{x}_i),\ j=1,N\right\}$, via a vector of
coefficients $\myvec{a}$.  In matrix form, the relation can be written
\begin{equation}
  \myvec\gamma = \myvec\Psi\myvec{a}
\end{equation}
where the columns of the matrix $\myvec\Psi$ are the basis vectors 
$\myvec\Psi_i$.  Orthonormality is given by the condition 
$\myvec\Psi_i^\dagger\myvec\Psi_j = \delta_{ij}$, so that the coefficients
can be determined by
\begin{equation}
  \myvec{a} = \myvec{\Psi}^\dagger\myvec{\gamma}
\end{equation}
A KL decomposition is optimal in the sense that it seeks basis 
functions for which the 
coefficients are statistically orthogonal;\footnote{Note that statistical
orthogonality of coefficients is conceptually distinct from the 
geometric orthogonality of the basis functions themselves; 
see \citet{Vogeley96} for a discussion of this property.}
that is, they satisfy
\begin{equation}
  \langle a_i^* a_j \rangle = \langle a_i^2 \rangle \delta_{ij}
\end{equation}
where angled braces $\langle\cdots\rangle$ denote averaging over all 
realizations.  This definition leads to several important properties
\citep[see][for a thorough discussion \& derivation]{Vogeley96}:
\begin{enumerate}
\item \textbf{KL as an Eigenvalue Problem:} 
  Defining the correlation matrix 
  $\myvec{\xi}_{ij} = \langle \gamma_i\gamma_j^*\rangle$, 
  it can be shown that the KL vectors $\myvec{\Psi}_i$ are eigenvectors 
  of $\myvec{\xi}$ with eigenvalues $\lambda_i = \langle a_i^2\rangle$.
  For clarity, we'll order the eigenbasis such that 
  $\lambda_i \ge \lambda_{i+1}\ \forall\ i\in(1,N-1)$.  We define the
  diagonal matrix of eigenvalues $\mymat{\Lambda}$, such that
  $\mymat{\Lambda}_{ij} = \lambda_i\delta_{ij}$
  and write the eigenvalue decomposition in compact form:
  \begin{equation}
    \mymat{\xi} = \mymat{\Psi}\mymat{\Lambda}\mymat{\Psi}^\dagger
  \end{equation}

\item \textbf{KL as a Ranking of Signal-to-Noise}
  It can be shown that KL vectors of a whitened covariance matrix (see
  Section~\ref{Adding_Noise})
  diagonalize both the signal and the noise of the problem, with the
  signal-to-noise ratio proportional to the eigenvalue.  This is
  why KL modes are often called ``Signal-to-noise eigenmodes''.

\item \textbf{KL as an Optimal Low-dimensional Representation:}
  An important consequence of the signal-to-noise properties of KL modes  
  is that the optimal rank-$n$ representation of the data is 
  contained in the KL vectors corresponding to the $n$ largest eigenvalues:
  that is,
  \begin{equation}
    \label{eq_truncation}
    \myvec{\hat\gamma}^{(n)}
    \equiv \sum_{i=1}^{n<N} a_i\myvec{\Psi}_i
  \end{equation}
  minimizes the reconstruction error between $\myvec{\hat\gamma}^{(n)}$ and 
  $\myvec\gamma$ for reconstructions using $n$ orthogonal basis vectors.
  This is the theoretical basis of Principal Component Analysis (sometimes
  called Discrete KL), and leads to a common application of KL 
  decomposition: filtration of noisy signals.  For notational compactness,
  we will define the truncated eigenbasis $\mymat{\Psi}_{(n)}$ and truncated
  vector of coefficients $\myvec{a}_{(n)}$ such that 
  Equation~\ref{eq_truncation} can be written in matrix form:
  $\myvec{\hat\gamma}^{(n)} = \mymat{\Psi}_{(n)}\myvec{a}_{(n)}$.
\end{enumerate}
 
\subsection{KL in the Presence of Noise}
\label{Adding_Noise}
When noise is present in the data, the above properties do not 
necessarily hold.
To satisfy the statistical orthogonality of the KL coefficients $\myvec{a}$ 
and the resulting signal-to-noise properties of the KL eigenmodes, 
it is essential that the noise in the covariance matrix be ``white'': 
that is, $\Noise_\gamma \equiv 
\langle \myvec{n}_\gamma\myvec{n}_\gamma^\dagger \rangle \propto \mymat{I}$.  
This can be accomplished through a judicious
choice of binning, or by rescaling the covariance with a whitening 
transformation.  We take the latter approach here.

Defining the noise covariance matrix $\Noise_\gamma$ as above,
the whitened covariance matrix can be written 
$\myvec{\xi}_W = \Noise_\gamma^{-1/2} \myvec{\xi} \Noise_\gamma^{-1/2}$.  Then 
the whitened KL modes become
$\myvec{\Psi}_W\myvec{\Lambda}_W\myvec{\Psi}_W^\dagger \equiv \myvec{\xi}_W$.
The coefficients $\myvec{a}_W$ are calculated from the noise-weighted signal,
that is
\begin{equation}
  \myvec{a}_W = \myvec{\Psi}_W^\dagger\Noise_\gamma^{-1/2}
  (\myvec{\gamma}+\myvec{n}_\gamma)
\end{equation}
For the whitened KL modes, if signal and noise are uncorrelated, this leads to 
$\langle\myvec{a}_W\myvec{a}_W^\dagger\rangle = \myvec{\Lambda}_W + \myvec{I}$:
that is, the coefficients $\myvec{a}_W$ are statistically orthogonal.
For the remainder of this work, we will drop the subscript ``$_W$'' and assume
all quantities to be those associated with the whitened covariance.

\subsection{Computing the Shear Correlation Matrix}
\label{Shear_Correlation}
The KL reconstruction of shear requires knowledge of the
form of the pixel-to-pixel correlation matrix $\myvec\xi$.  
In many applications of KL
\citep[e.g.~analysis of galaxy spectra,][]{Connolly95} 
this correlation matrix is
determined empirically from many realizations of the data (i.e.~the set
of observed spectra).  In the case of weak lensing shear, we generally
don't have many realizations of the data, so this approach is not
tenable.  Instead, we compute this correlation matrix analytically.  The 
correlation of the cosmic shear signal between two regions of the sky
$A_i$ and $A_j$ is given by
\begin{eqnarray}
  \label{xi_analytic}
  \myvec{\xi}_{ij} 
  &=& \langle\gamma_i\gamma_j^*\rangle + 
  \langle n_in_j^*\rangle\nonumber\\
  &=& \left[\int_{A_i}d^2x_i\int_{A_j}d^2x_j 
    \xi_+(|\myvec{x_i}-\myvec{x_j}|)\right]
  + \delta_{ij}\frac{\sigma_\epsilon^2}{\bar{n}}
\end{eqnarray}
where $\sigma_\epsilon$ is the intrinsic shape noise (typically assumed to 
be $\sim 0.3$), $\bar{n}$ is the average galaxy count per pixel, and 
$\xi_+(\theta)$ is the ``+'' shear correlation function \citep{Schneider02}. 
$\xi_+(\theta)$ can be expressed as an integral over the shear power spectrum:
\begin{equation}
  \label{xi_plus_def}
  \xi_+(\theta) 
  = \frac{1}{2\pi} \int_0^\infty d\ell\ \ell P_\gamma(\ell) J_0(\ell\theta)
\end{equation}
where $J_0$ is the zeroth-order Bessel function of the first kind.  The 
shear power spectrum $P_\gamma(\ell)$ can be expressed as an 
appropriately weighted line-of-sight integral over the 3D mass power 
spectrum \citep[see, e.g.][]{Takada04}:
\begin{equation}
  \label{P_gamma}
  P_\gamma(\ell) = \int_0^{\chi_s}d\chi W^2(\chi)\chi^{-2}
  P_\delta\left(k=\frac{\ell}{\chi};z(\chi)\right)
\end{equation}
Here $\chi$ is the comoving distance, $\chi_s$ is the distance to the
source, and $W(\chi)$ is the lensing weight function,
\begin{equation}
  \label{Lensing_Weight}
  W(\chi) = \frac{3\Omega_{m,0}H_0^2}{2a(\chi)}\frac{\chi}{\bar{n}_g}
  \int_{z(\chi)}^{z(\chi_s)}dz\ n(z) \frac{\chi(z)-\chi}{\chi(z)}
\end{equation}
where $n(z)$ is the redshift distribution of galaxies.  We assume a
DES-like survey, where $n(z)$ has the approximate form
\begin{equation}
  \label{Number_Distribution}
  n(z) \propto z^2 \exp[-(z/z_0)^{1.5}]
\end{equation}
with $z_0 = 0.5$, where $n(z)$ is normalized to the observed galaxy density
$\bar{n}_g = 20\ {\rm arcmin}^{-2}$.

The 3D mass power spectrum $P_\delta(k,z)$ in Equation~\ref{P_gamma}
can be computed theoretically.  
In this work we compute $P_\delta(k,z)$ using the halo model of 
\citet{Smith03}, and compute the correlation matrix $\myvec\xi$ using 
Equations~\ref{xi_analytic}-\ref{Number_Distribution}.
When computing the double integral of Equation~\ref{xi_analytic},
we calculate the integral in two separate regimes:
for large separations ($\theta > 20$ arcmin), 
we assume $\xi_+(\theta)$ doesn't change appreciably over the area 
of the pixels, so that only a single evaluation of
the $\chi_+(\theta)$ is necessary for each pixel pair.  
For smaller separations, 
this approximation is insufficient, and we evaluate $\myvec\xi_{ij}$ using
a Monte-Carlo integration scheme.  Having calculated the 
theoretical correlation matrix $\myvec\xi$ for a given field, 
we compute the KL basis directly using an eigenvalue decomposition.

\subsection{Which Shear Correlation?}
\label{WhichCorrelation}
Above we note that the correlation matrix of the measured shear 
can be expressed
in terms of the ``+'' correlation function, $\xi_+(\theta)$.  This is not
the only option for measurement of shear correlations 
\citep[see, e.g.][]{Schneider02}.  So why use $\xi_+(\theta)$ rather 
than $\xi_-(\theta)$?  The answer lies in the KL formalism itself.  The KL
basis of a quantity $\myvec\gamma$ is constructed via its correlation
$\langle\myvec\gamma \myvec\gamma^\dagger\rangle$.  Because of the complex
conjugation involved in this expression, the only relevant correlation 
function for KL is $\xi_+(\theta)$ \textit{by definition}.  Nevertheless,
one could object that by neglecting $\xi_-$, 
KL under-utilizes the theoretical information available 
about the correlations of cosmic shear. However, in the absence of noise, 
the two correlation functions contain identical information: 
either function can be determined from the other.  
In this sense, the above KL formalism
uses all the shear correlation information that is available.

One curious aspect of this formalism is that the theoretical covariance
matrix and associated eigenmodes are real-valued, while the shear we are
trying to reconstruct is complex-valued.  This can be traced
to the computation of the shear correlation:
\begin{equation}
  \xi_+ \equiv \langle\gamma \gamma^*\rangle
  = \langle \gamma_t\gamma_t\rangle + \langle \gamma_\times\gamma_\times\rangle
  + i[\langle\gamma_t\gamma_\times\rangle - \langle\gamma_\times\gamma_t\rangle]
\end{equation}
By symmetry, the imaginary part of this expression
is zero. At first glance, this might seem a bit strange:
how can a complex-valued data vector be reconstructed from a
real-valued orthogonal basis?  The answer lies in the complex KL coefficients
$a_i$: though each KL mode contributes only a single phase across the field
(given by the phase of the associated $a_i$), the reconstruction has a 
plurality of phases due to the varying magnitudes of the contributions 
at each pixel (given by the elements of each basis vector $\myvec{\Psi}_i$).

An important consequence of this observation is that the KL modes 
themselves are not sensitive by construction to the E-mode (curl-free) 
and B-mode (divergence-free) components of the shear field. As we will
show below, however, the signal-to-noise properties of KL modes lead to 
some degree of sensitivity to the E and B-mode information in a given
shear field (See Section~\ref{Discussion}).

\subsection{Interpolation using KL Modes}
\label{KL_Interpolation}
Shear catalogs, in general, are an incomplete and inhomogeneous
tracer of the underlying shear field, and some regions of the field may 
contain no shear information.  This sparsity of data poses a problem,
because the KL modes are no longer orthogonal over the incomplete field.
\citet{Connolly99} demonstrated how this missing-information problem can be 
addressed for KL decompositions of galaxy
spectra. We will summarize their results here.  First we define the weight 
function $w(\myvec{x}_i)$.
The weight function can be defined in one of two ways: 
a binary weighting convention where
$w(\myvec{x}_i)=0$ in masked pixels and $1$ elsewhere, or a continuous 
weighting convention where $w(\myvec{x}_i)$ scales inversely with the noise 
$[\Noise_\gamma]_{ii}$.
The binary weighting convention treats the noise is part of the data, 
and so the measurements should be whitened as outlined in 
Section~\ref{Adding_Noise}.  
The continuous weighting convention assumes the noise is part of the mask, 
so data and noise are not whitened.
We find that the two approaches lead to qualitatively similar results, 
and choose to use the binary weighting convention for the simplicity of 
comparing masked and unmasked cases.

Let $\myvec{\gamma}^o$ be the observed data vector, 
which is unconstrained where
$w(\myvec{x}_i)=0$.  Then we can obtain the KL coefficients $a_i$ by 
minimizing the reconstruction error of the whitened data
\begin{equation}
  \label{chi2_min}
  \chi^2 = ( \Noise_\gamma^{-1/2}\myvec{\gamma}^o
  - \myvec{\Psi}_{(n)}\myvec{a}_{(n)} )^\dagger 
  \myvec{W}
  ( \Noise_\gamma^{-1/2}\myvec{\gamma}^o
  - \myvec{\Psi}_{(n)}\myvec{a}_{(n)} )
\end{equation}
where we have defined the diagonal weight matrix 
$\myvec{W}_{ij} = w(\myvec{x}_i) \delta_{ij}$.
Minimizing Equation~\ref{chi2_min} with respect to $\myvec{a}$ leads to
the optimal estimator $\myvec{\hat{a}}$, which can be expressed
\begin{equation}
 \myvec{\hat{a}}_{(n)} = 
 \mymat{M}_{(n)}^{-1} 
 \myvec{\Psi}_{(n)}^\dagger \myvec{W} \Noise_\gamma^{-1/2}\myvec{\gamma}^o
\end{equation}
Where we have defined the mask convolution matrix 
$\mymat{M}_{(n)} \equiv \mymat{\Psi}_{(n)}^\dagger\mymat{W}\mymat{\Psi}_{(n)}$.
These coefficients $\myvec{\hat{a}}_{(n)}$
can then be used to construct an estimator for the unmasked shear field:
\begin{equation}
  \label{shear_recons}
  \myvec{\hat{\gamma}}^{(n)} = \Noise_\gamma^{1/2}\myvec{\Psi}_{(n)}\myvec{\hat{a}}_{(n)}
\end{equation}
In cases where the mask convolution matrix $\mymat{M}_{(n)}$  
is singular or nearly singular, the estimator in Equation~\ref{shear_recons}
can contain unrealistically large values within the reconstruction 
$\myvec{\hat{\gamma}}^{(n)}$.
This can be addressed either by reducing $n$, or by 
adding a penalty function to the right side of Equation~\ref{chi2_min}.  
One convenient form of this penalty is the generalized
Wiener filter \citep[see][]{Tegmark97}, which penalizes results which 
deviate from the expected correlation matrix.  Because the correlation
matrix has already been computed when determining the KL modes, 
this filter requires very little extra computation.
With Wiener filtering, Equation~\ref{chi2_min} becomes
\begin{eqnarray}
  \label{chi2_min_WF}
  \chi^2 & = & ( \Noise_\gamma^{-1/2}\myvec{\gamma}^o - \myvec{\Psi}_{(n)}\myvec{a}_{(n)} )^\dagger 
  \myvec{W} 
  (\Noise_\gamma^{-1/2}\myvec{\gamma}^o - \myvec{\Psi}_{(n)}\myvec{a}_{(n)} )
  \nonumber\\
  && + \alpha\ \myvec{a}_{(n)}^\dagger\myvec{C}_{a(n)}^{-1}\myvec{a}_{(n)}
\end{eqnarray}
where $\myvec{C}_{a(n)}\equiv\langle 
\myvec{a}_{(n)}\myvec{a}_{(n)}^\dagger\rangle$ 
and $\alpha$ is a tuning parameter which lies in the range $0\le\alpha\le 1$. 
Note that for $\alpha=0$, the result is the same as in the unfiltered case.
Minimizing Equation~\ref{chi2_min_WF} with respect to $\myvec{a}$ gives the
filtered estimator
\begin{equation}
  \label{a_WF}
  \myvec{\hat{a}}_{(n,\alpha)} = 
  \mymat{M}_{(n,\alpha)}^{-1} 
  \myvec{\Psi}_{(n)}^\dagger \myvec{W} \Noise_\gamma^{-1/2}\myvec{\gamma}^o
\end{equation}
where we have defined $\mymat{M}_{(n,\alpha)} 
\equiv [\myvec{\Psi}_{(n)}^\dagger\myvec{W}\myvec{\Psi}_{(n)} 
  + \alpha\myvec{\Lambda}_{(n)}^{-1}]$, and
$\myvec{\Lambda}_{(n)}$ is the truncated diagonal matrix of 
eigenvalues associated with $\myvec{\Psi}_{(n)}$.

\begin{figure*}
 \centering
 \plotone{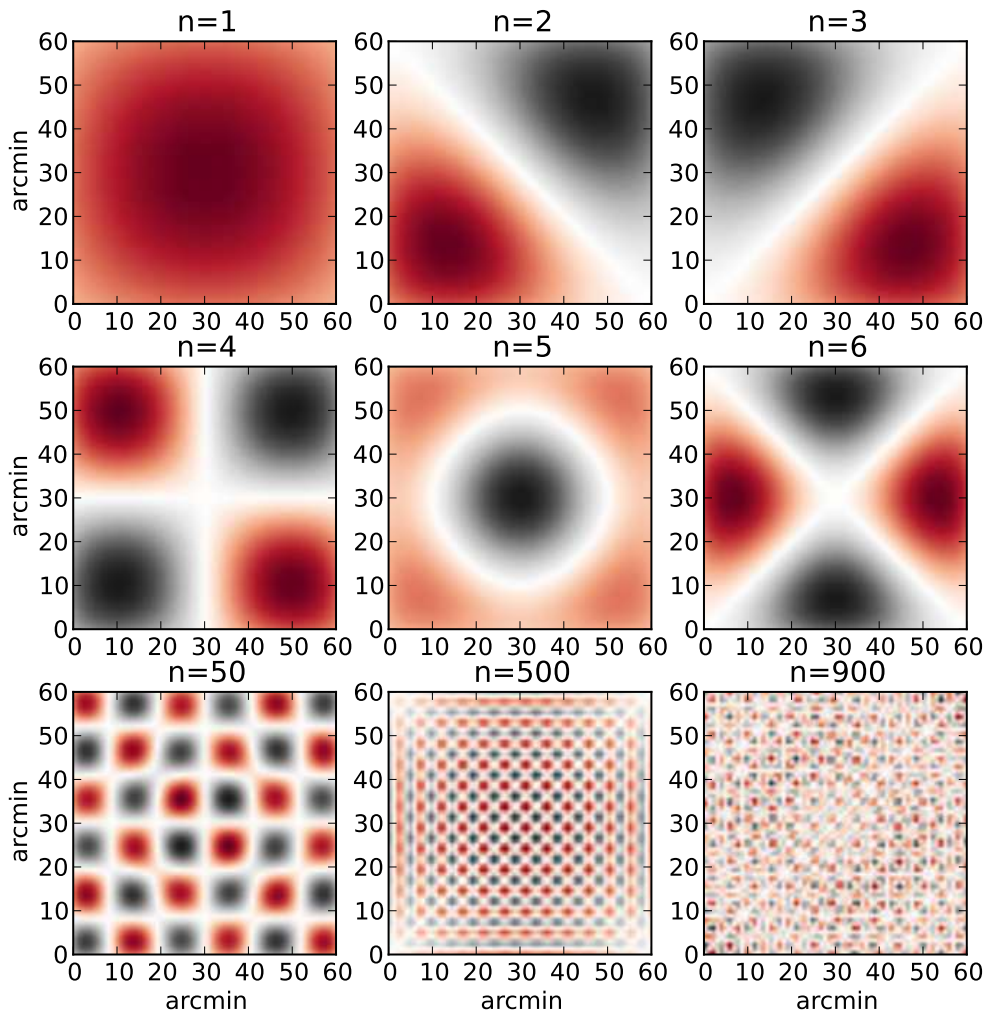} 
 \caption{
   A sample of nine of the 4096 KL eigenmodes 
   of a $1^\circ\times 1^\circ$ patch of the sky partitioned into
   $64\times 64$ pixels.  Black is positive, red is negative, and each mode
   has unit norm. The modes are calculated from the theoretical
   shear correlation function (see Section~\ref{Shear_Correlation}).  
   As a consequence of the isotropy of the cosmic shear field,
   the covariance matrix -- and thus the associated eigenmodes --
   are purely real (see Section~\ref{Testing_Shear_KL}).
   \label{fig_KL_modes} }
\end{figure*} 

\section{Testing KL Reconstructions}
\label{Testing_Reconstruction}
In this section we show results of the KL analysis of shear fields for
a sample geometry.  In Section~\ref{Testing_Shear_KL} we discuss the general
properties of shear KL modes for unmasked fields, while in
Section~\ref{Testing_Interpolation} we discuss KL shear reconstruction 
in the presence of masking. 

\subsection{KL Decomposition of a Single Field}
\label{Testing_Shear_KL}
To demonstrate the KL decomposition of a shear field, we assume a square field
of size $1^\circ\times 1^\circ$, divided into $64\times 64$ pixels.  We assume
a source galaxy density of $20\ \mathrm{arcmin}^{-2}$ -- appropriate for a
ground-based survey such as DES -- and calculate the
KL basis following the method outlined in Section~\ref{KL_Formalism}.
For the computation of the nonlinear matter power spectrum, we assume a flat 
$\Lambda$CDM cosmology with $\Omega_M=0.27$ at the present day, with
the power spectrum normalization given by $\sigma_8=0.81$.

\begin{figure*}
 \centering
 \plotone{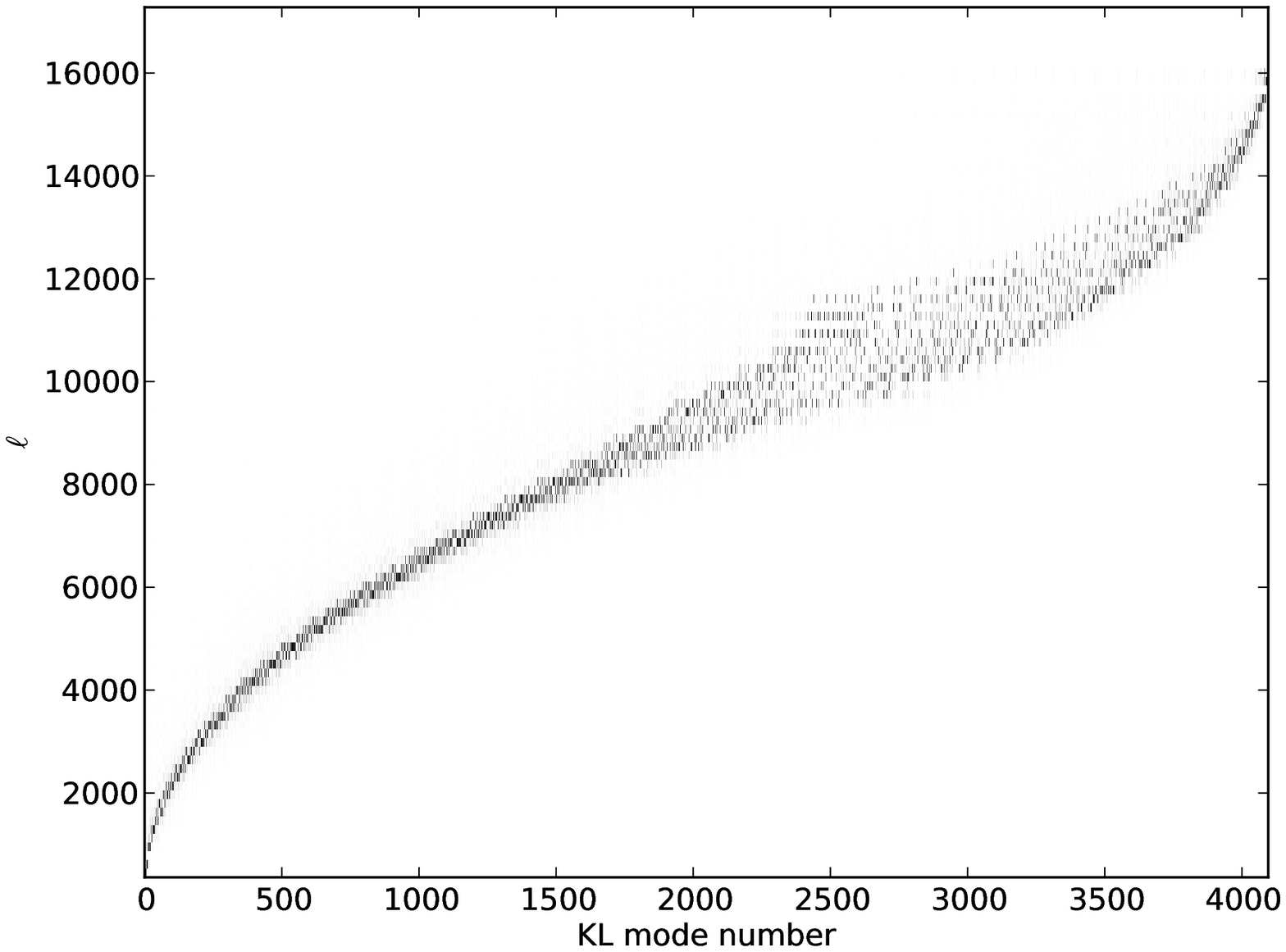}
 \caption{The normalized power spectrum of each KL mode.  For constant
   mode number, the figure represents a histogram of the power in that KL
   mode, normalized to a constant total power.  KL modes represent a 
   linear combination of Fourier modes, so that the power in each KL 
   mode is spread over a range of $\ell$ values.  Nevertheless,
   the general trend is clear: larger mode numbers are associated with
   larger wave numbers, and thus smaller length scales.
   \label{fig_bandpower} }
\end{figure*}

Figure~\ref{fig_KL_modes} shows a selection of
nine of the 4096 shear eigenmodes within
this framework.  The KL modes are reminiscent of 2D Fourier modes, with
higher-order modes probing progressively smaller length scales.  
This characteristic length scale of the eigenmodes
can be seen quantitatively in Figure~\ref{fig_bandpower}.  
Here we have computed the rotationally averaged
power spectrum $C_\ell$ for each individual Fourier mode, and plotted the
power vertically as a density plot for each mode number.  
Because the KL modes are
not precisely equivalent to the 2D Fourier modes, each contains power at
a range of values in $\ell$.  But the overall trend is clear: larger modes
probe smaller length scales, and the modes are very close to Fourier in
nature.

\begin{figure*}
 \centering
 \plotone{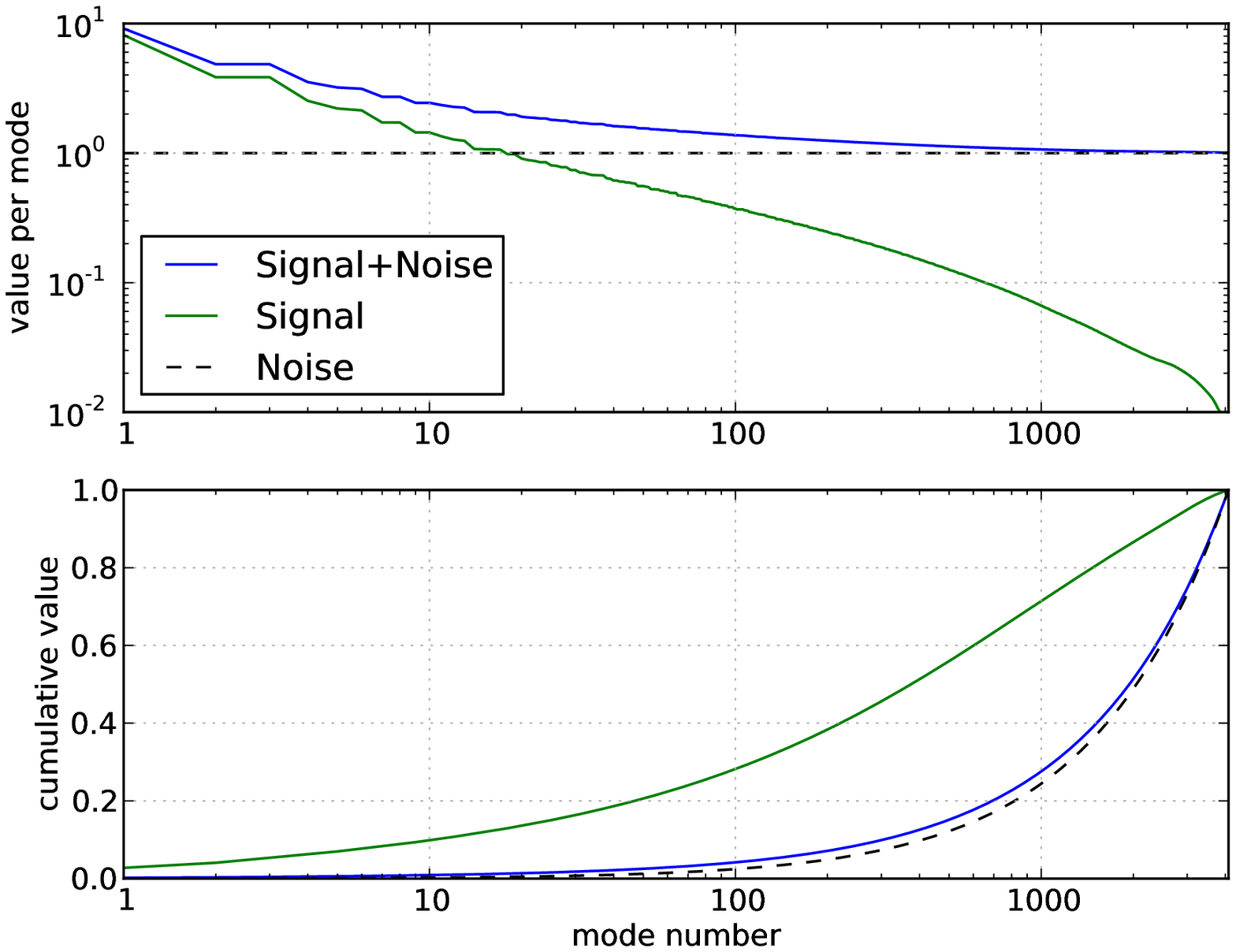}
 \caption{
   The eigenvalues associated with the eigenmodes discussed in
   Figure~\ref{fig_KL_modes}.  By construction, the eigenvalue is 
   proportional to the sum of signal and
   noise within each mode.  The upper figure shows the value per mode,
   while the lower figure shows the normalized cumulative value.
   The stepped-pattern evident in the upper panel is due to the presence
   of degenerate eigenmodes which have identical eigenvalues
   (e.g. modes $n=2$ and $n=3$, related by parity as 
   evident in Figure~\ref{fig_KL_modes}).
   Because the eigenmodes are computed from a whitened covariance matrix 
   (see Section~\ref{Adding_Noise}), the noise contribution within each
   mode is equal to 1.  Subtracting this contribution leads to the plot of
   signal only: this shows that the signal-to-noise ratio is above unity 
   only for the first $17$ modes.  Still, as seen in the lower panel,
   higher modes are required: the first $17$ modes account for only $12\%$
   of the signal on average.
   To recover $70\%$ of the signal in a particular reconstruction requires
   about 1000 modes.  At this point, each additional mode has a signal-to-noise
   ratio of less than $0.1$.  Such a small signal-to-noise ratio is a
   well-known aspect of cosmic shear studies.
   \label{fig_eigenvalues} }
\end{figure*}

As noted in Section~\ref{KL_Intro}, one useful quality of a KL decomposition
is its diagonalization of the signal and noise of the problem.
To explore this property, we plot in the upper panel of 
Figure~\ref{fig_eigenvalues} the eigenvalue profile 
of these KL modes. By construction, higher-order modes 
have smaller KL eigenvalues.  What is more,
because the noise in the covariance matrix is whitened 
(see Section~\ref{Adding_Noise}), the expectation of the noise covariance
within each mode is equal to 1.  Subtracting this noise from each eigenvalue 
gives the expectation value of the signal-to-noise ratio: 
thus we see that the expected
signal-to-noise ratio of the eigenmodes is above unity only for the first
17 of the 4096 modes.

At first glance, this may seem to imply that only the first 17 or so modes
are useful in a reconstruction.  On the contrary: as seen in the lower panel
of Figure~\ref{fig_eigenvalues}, these first 17 modes contain only a
small fraction of the total information in the shear field (This is not 
an unexpected result: cosmic shear measurements have 
notoriously low signal-to-noise ratios!)
About 900 modes are needed to preserve an average of 70\% of the total signal, 
and at this level, each additional mode has a signal-to-noise ratio 
of below $1/10$.  The noisy input shear field can be exactly recovered
by using all 4096 modes: in this case, though, the final few modes 
contribute two orders-of-magnitude more noise than signal.

\begin{figure*}
 \centering
 \plotone{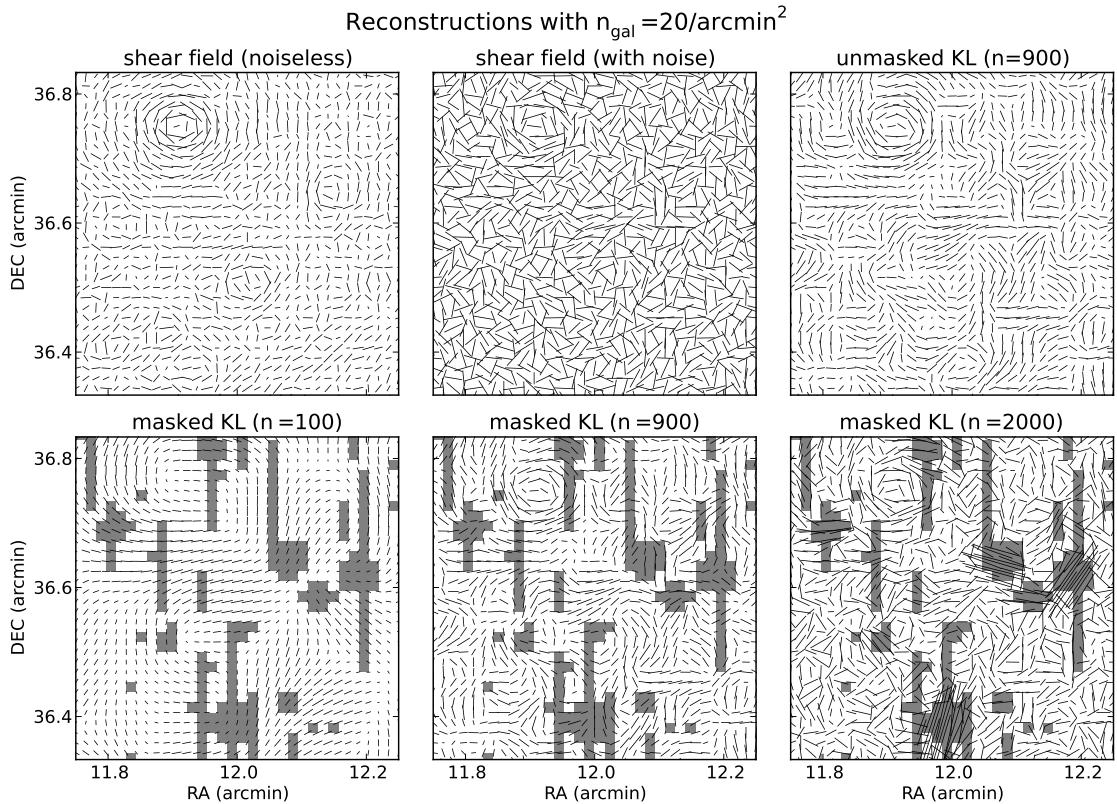}
 \caption{
   This figure illustrates the reconstruction of a small patch of masked
   shear from simulated shear catalog.
   \textit{upper panels:} 
   The underlying noiseless shear signal \textit{(left)},
   the observed, noisy shear signal \textit{(middle)},
   and the unmasked reconstruction with 900 modes and $\alpha=0.15$.
   The amplitude of the noise is calculated using an intrinsic ellipticity 
   $\sigma_\epsilon = 0.3$, with an average number density of 
   $n_{gal}=20\ \mathrm{arcmin}^{-2}$.  The large peak in the upper 
   portion of the figure is well-recovered by the KL reconstruction.
   \textit{lower panels:} The KL reconstruction of the shear in the
   presence of 20\% masking, with increasing number of modes $n$.
   The mask is represented by the shaded regions in panels: 
   within these regions, the value of the shear is
   recovered through KL interpolation (see Section~\ref{KL_Interpolation}).
   We see in this progression the effect of the KL cutoff choice:
   using too few modes leads to loss of information, 
   while using too many modes leads to over-fitting within the masked regions 
   (See Appendix~\ref{Choosing_Params} for a discussion of the choice of
   number of modes).
   \label{fig_reconstruction} }
\end{figure*}

\begin{figure*}
 \centering
 \plotone{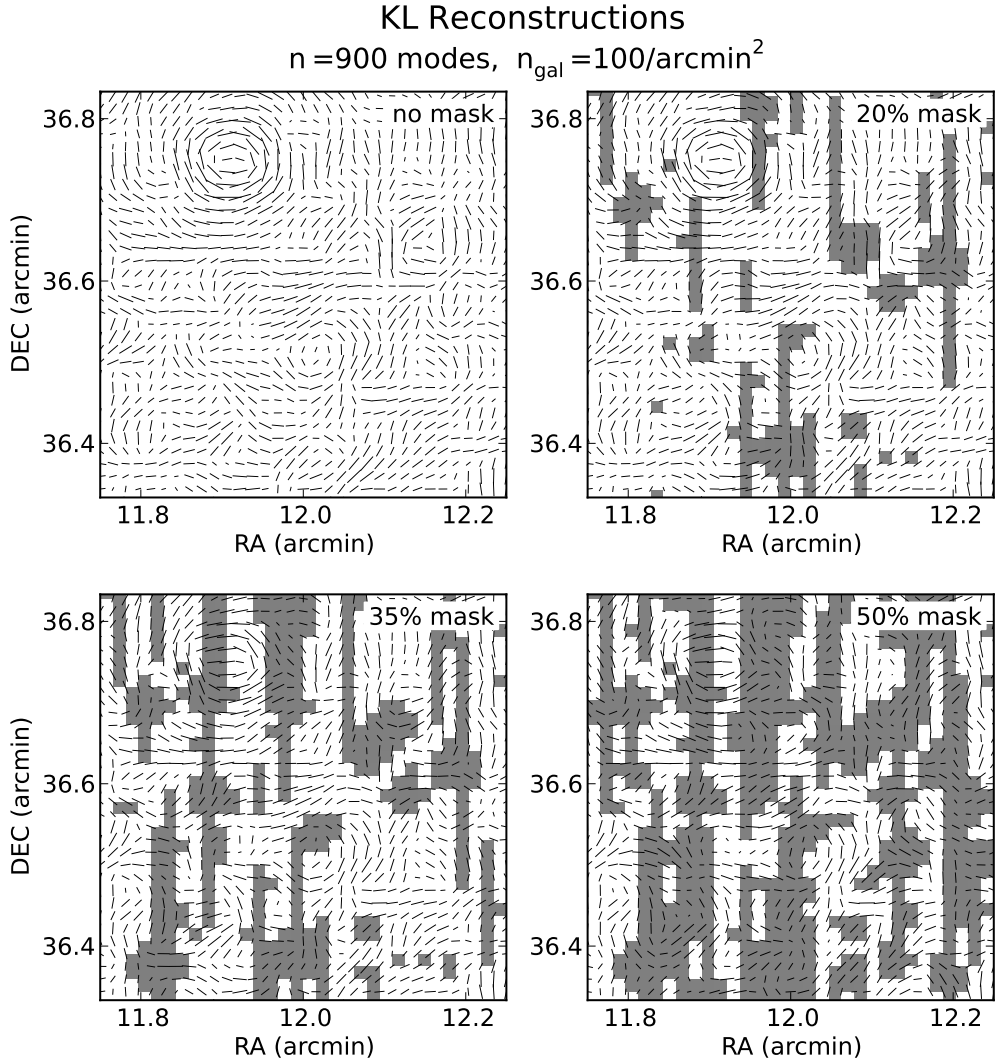}
 \caption{
   Here we show the same field as in Figure~\ref{fig_reconstruction}, 
   reconstructed using $n=900$ modes, with increasing levels of mask
   coverage.  The density of source galaxies has been increased to 
   100 arcmin$^{-2}$, typical of a space-based weak lensing survey.  
   At this noise
   level, smaller halos can be detected within the unmasked KL
   reconstruction (upper-left panel).  Even at a 50\% masking level, 
   the large peak at (RA,DEC) = (11.9,36.75) is adequately recovered.
   \label{fig_reconstruction_2} }
\end{figure*}

\subsection{Testing KL Interpolation}
\label{Testing_Interpolation}
To test this KL interpolation technique, we use simulated shear 
catalogs\footnote{The simulated shear catalogs were kindly made available 
to us by R. Wechsler, M Busha, and M. Becker.}. 
These catalogs contain 220 square degrees of simulated shear maps, computed 
using a ray-tracing grid through a cosmological N-body simulation of the 
standard $\Lambda$-CDM model. The shear signal is computed at the locations 
of background galaxies with a median redshift of about 0.7.  Galaxies are 
incorporated in the simulation using the ADDGALS algorithm 
\citep[][Wechsler \textit{et al.} in preperation]{Wechsler04}, 
tuned to the expected observational characteristics of the DES mission. 

We pixelize this shear field using the same pixel size as above: 
$64\times 64$ pixels per square degree.
To perform the KL procedure on the full field with this angular 
resolution would lead to a data vector containing over $10^6$ elements, 
and an associated covariance matrix containing $10^{12}$ entries.  
A full eigenvalue decomposition of such a matrix is 
computationally infeasible, so we reconstruct the field in 
$1^\circ\times 1^\circ$ tiles, each $64\times 64$ pixels in size.  To reduce
edge effects between these tiles, we use only the central 
$0.5^\circ\times 0.5^\circ$ region of each, so that covering the 300 square 
degree field requires 1200 tiles. 

In order to generate a realistic mask over the field area, 
we follow the procedure outlined in
\citet{Hikage10} which generates pixel-level masks characteristic of  
point-sources, saturation spikes, and bad CCD regions.  We tune the mask
so that 20\% of the shear pixels have no data.  The geometry of the mask
over a representative patch of the field can be seen in the lower 
panels of Figure~\ref{fig_reconstruction}, where we also show the result of
the KL interpolation using 900 out of 4096 modes, with $\alpha = 0.15$
(For a discussion of these parameter choices, see 
Appendix~\ref{Choosing_Params}).

The upper panels of Figure~\ref{fig_reconstruction} give a qualitative view
of the difficulty of cosmic shear measurements.  The top left panel shows the
noiseless shear across the field, while the top right panel shows the
shear with shape noise for a DES-type survey ($\sigma_\epsilon=0.3$, 
$\bar{n}_{gal} = 20\ \mathrm{arcmin}^{-2}$).  To the eye, the signal seems
entirely washed out by the noise.  Nevertheless, the shear signal is there,
and can be fairly well-recovered using the first 900 KL modes 
(middle-left panel). For masked data, we must resort to the 
techniques of Section~\ref{KL_Interpolation} to fill-in the missing data.
The middle-right panel shows this reconstruction, with gray shaded
regions representing the masked area.  A visual comparison of the masked 
and unmasked $n=900$ panels of Figure~\ref{fig_reconstruction} confirms
qualitatively that the KL interpolation is performing as desired.  This is
especially apparent near the large cluster located at (RA,DEC)=(11.9,36.7).  
The remaining two lower panels of Figure~\ref{fig_reconstruction} 
show cases of over-fitting and under-fitting
of the shear data. If too few KL modes are used, the structure of the input 
shear field is lost.  If too many KL modes are used, the masked regions 
are over-fit,
causing the interpolated shear values to become unnaturally large.  This
observation suggests one rubric by which the ideal number of modes can be
chosen; see the discussion in Appendix~\ref{Choosing_Params}.

It is interesting to explore the limits of this interpolation algorithm.
Figure \ref{fig_reconstruction_2} shows the KL reconstruction with
increasing masked fractions, using a noise level typical of space-based
lensing surveys $(n_{gal}=100{\rm arcmin}^{-2})$  Though the quality of
the reconstruction understandably degrades, the lower panels show that
large features can be recovered even with up to 50\% of the pixels masked.


For a quantitative analysis of the effectiveness of the KL interpolation 
in convergence mapping, and the potential biases it introduces, 
a large-scale statistical measure is most appropriate.
In the following sections, we test the utility of this KL interpolation scheme
within the framework of shear peak statistics.

\section{Shear Peak Statistics}
\label{Shear_Peaks}
It has long been recognized that much useful cosmological information 
can be deduced from the masses and spatial distribution of galaxy clusters 
\citep[e.g.][]{Press74}.  
Galaxy clusters are the largest gravitationally bound objects in the universe,
and as such are exponentially sensitive to cosmological parameters
\citep{White93}.  The spatial distribution of clusters and redshift evolution
of their abundance and clustering is sensitive to both geometrical effects
of cosmology, as well as growth of structure.  Because of this, cluster
catalogs can be used to derive constraints on many interesting 
cosmological quantities, including the matter density $\Omega_M$ 
and power spectrum normalization $\sigma_8$ \citep{Lin03}, 
the density and possible evolution of dark energy \citep{Linder03,
 Vikhlinin09b},
primordial non-gaussianities \citep{Matarrese00,Grossi07}, 
and the baryon mass fraction \citep{Lin03,Giodini09}.

Various methods have been developed to measure the mass and 
spatial distribution of galaxy clusters, and each are subject to 
their own difficult astrophysical and observational biases.
They fall into four broad categories: optical or infrared richness, 
X-ray luminosity and surface brightness, 
Sunyaev-Zeldovich decrement, and weak lensing shear.

While it was long thought that weak gravitational lensing studies would lead
to robust, purely mass-selected cluster surveys, it has since become 
clear that shape noise and projection effects limit the usefulness 
of weak lensing in determining the 3D cluster mass function 
\citep{Hamana04,Hennawi05,Mandelbaum10,Vanderplas11}.  
The shear observed in weak lensing is non-locally related to the convergence,
a measure of \textit{projected} mass along the line of sight.  
The difficulty in deconvolving the correlated and uncorrelated projections
in this quantity leads to difficulties in relating these projected peak
heights to the masses of the underlying clusters in three dimensions.
Recent work has shown, however, that this difficulty 
in relating the observed quantity 
to theory may be overcome through the use of statistics of
the projected density itself.

\citet{Marian09,Marian10} first explored the extent to which 2D 
projections of the
3D mass field trace cosmology.  They found, rather surprisingly, that the
statistics of the projected peaks closely trace the statistics of the 3D
peak distribution: in N-body simulations, both scale 
with the \citet{Sheth99} analytic
scaling relations.  The same correlated projections which bias
cluster mass estimates contribute to a usable signal: 
statistics of projected mass alone can provide useful cosmological 
constraints, without the need for bias-prone conversions from peak 
height to cluster mass.

A host of other work has explored diverse aspects of these
shear peak statistics,
including tests of these methods with ensembles of N-body simulations
\citep{Wang09,Kratochvil10,Dietrich10},
the performance of various filtering functions
and peak detection statistics \citep{Pires09,Schmidt10,Kratochvil11},
exploration of the spatial correlation of noise with signal
within convergence maps \citep{Fan10},
and exploration of shear-peak constraints on primordial non-gaussianity
\citep{Maturi11}.  The literature has yet to converge
on the ideal mapping procedure: convergence maps, gaussian filters,
various matched filters, wavelet transforms, and more novel filters
are explored within the above references.  There is also variation in
how a ``peak'' is defined: simple local maxima, ``up-crossing'' criteria, 
fractional areas above a certain threshold, connected-component labeling, 
hierarchical methods, and Minkowski functionals are all shown to be useful.  
Despite diverse methodologies, all the above work confirms that there 
is useful cosmological information within the projected peak 
distribution of cosmic shear fields, and that this information adds to 
that obtained from 2-point statistics alone.

\subsection{Aperture Mass Peaks}
\label{Aperture_Mass}
Based on this consensus, we use shear peak statistics to explore
the possible bias induced by the KL interpolation method outlined above.  
We follow the aperture mass methodology of \citet{Dietrich10}:
The aperture mass magnitude at a point $\vec\theta_0$ is given by
\begin{equation}
  \label{Map_def}
  M_{\rm ap}(\vec{\theta_0})
  = \int_\Omega d^2\theta Q_{\rm NFW}(\vartheta=|\vec\theta-\vec\theta_0|) 
  \gamma_t(\vec\theta;\vec\theta_0)
\end{equation}
where $\gamma_t(\vec\theta;\vec\theta_0)$ is the component of the shear at 
location $\vec\theta$ tangential to the line $\vec\theta-\vec\theta_0$, 
and $Q_{\rm NFW}$ is the 
NFW-matched filter function defined in \citet{Schirmer07}:
\begin{equation}
  Q_{\rm NFW}(x;x_c) \propto \frac{1}{1+e^{6-150x}+e^{-47+50x}}
  \frac{\tanh(x/x_c)}{x/x_c}
\end{equation}
with $x = \vartheta/\vartheta_{\rm max}$ and $x_c$ a free parameter. We follow
\citet{Dietrich10} and set $\vartheta_{\rm max}=5.6\ \mathrm{arcmin}$ and
$x_c=0.15$.  The integral in Equation~\ref{Map_def} is over the whole sky, 
though the filter function $Q$ effectively cuts this off at a radius
$\vartheta_{\rm max}$.  
In the case of our pixelized shear field, the integral is converted to 
a discrete sum over all pixels, with
$\vartheta$ equal to the distance between the pixel centers:
\begin{equation}
  \label{Map_def_discrete}
  M_{\rm ap}(\vec{\theta}_i) 
  = \sum_j Q_{\rm NFW}(\vartheta_{ij}) 
  \gamma_t(\vec{\theta}_i;\vec{\theta}_j) 
\end{equation}
where we have defined $\vartheta_{ij} \equiv |\vec\theta_i-\vec\theta_j|$.

We can similarly compute the \textit{B-mode} aperture mass, by substituting
$\gamma_t \to \gamma_\times$ in Equations~\ref{Map_def}-\ref{Map_def_discrete}
\citep{Crittenden02}.  For pure gravitational
weak lensing with an unbiased shear estimator, the B-mode signal is 
expected to be negligible, though second-order effects such as 
source clustering and intrinsic 
alignments can cause contamination on small angular scales 
\citep{Crittenden02,Schneider02b}
These effects aside, the B-mode signal can be used as a rough estimate of the
systematic bias of a particular analysis method.

For our study, the aperture mass is calculated with the same resolution as the
shear pixelization: $64^2$ pixels per square degree.  A pixel is defined to be
a peak if its value is larger than that of the surrounding eight pixels:
a simple local maximum criterion. 

\begin{figure*}
 \centering
 \plotone{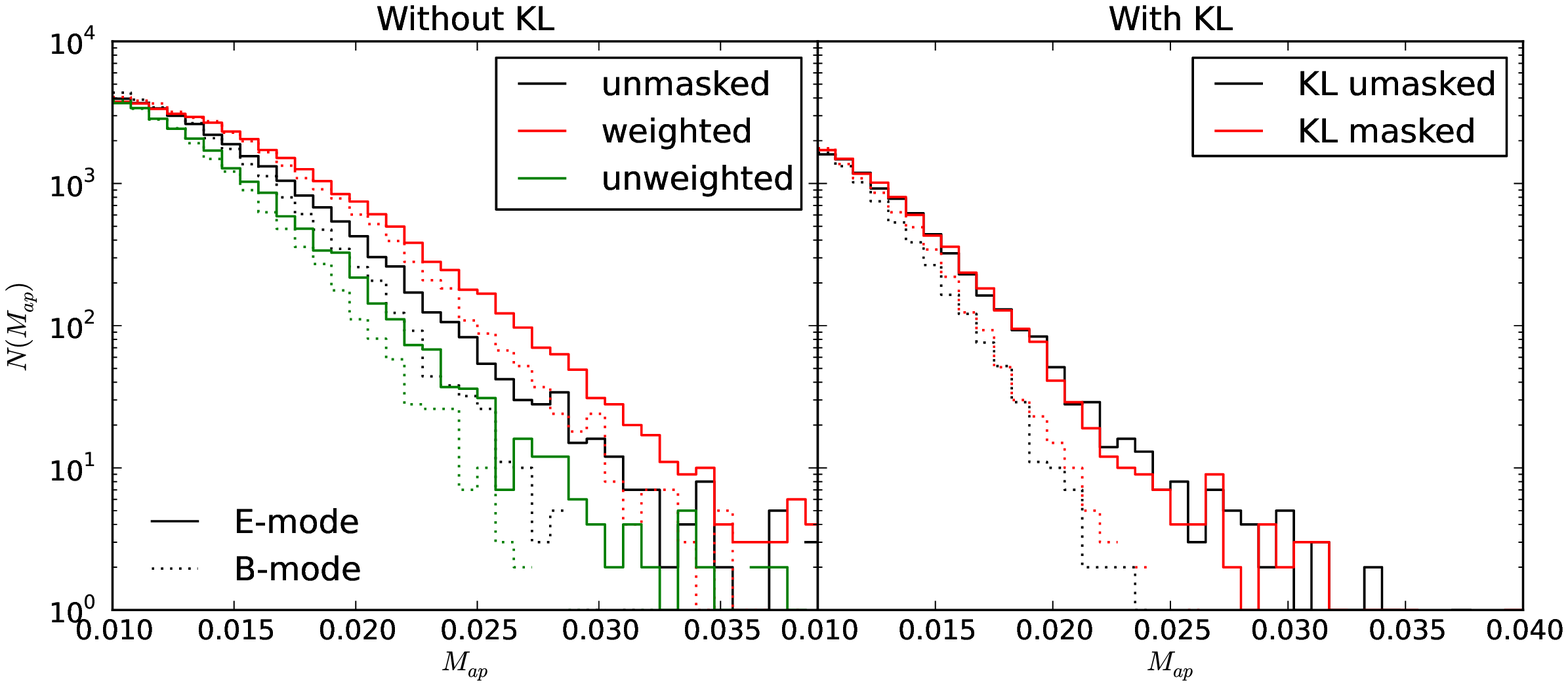}
 \caption{
   Comparison of the masked and unmasked peak distributions.
   \textit{left panel:} the peak distributions without the use of KL.
   The black line is the result with no masking, while the red and green
   lines show the two \naive methods of correcting for the mask (see
   Section~\ref{MaskingEffects}).
   \textit{right panel:}  the masked and unmasked peak distributions
   after applying KL. 
   Neither \naive method of mask-correction adequately recovers 
   the underlying peak distribution.  It is evident, however, that 
   the KL-based interpolation procedure recovers a mass map with 
   a similar peak distribution to the unmasked KL map.
   It should be noted that the unmasked peak distribution 
   (black line, left panel) is not identical to the unmasked peak distribution
   after application of KL (black line, right panel).  This difference is
   addressed in Figure~\ref{fig_num_peaks}.
   \label{fig_mask_nomask} 
 }
\end{figure*} 

\subsection{The Effects of Masking}
\label{MaskingEffects}
When a shear peak statistic is computed across a field with masked regions,
the masking leads to a bias in the peak height distribution
(see Section~\ref{Discussion} below).  Moreover, due
to the non-local form of the aperture mass statistic, a very large region is
affected: in our case, a single masked pixel biases the aperture mass 
measurement of an area of size
 $\pi(2\vartheta_{\rm max})^2 \approx 400\ {\rm arcmin}^2$.  
There are two \naive approaches one could use when measuring the 
aperture mass in this situation:

\begin{description}
  \item[Unweighted:] Here we simply set the shear value within each 
    masked pixel to zero, and apply Equation~\ref{Map_def_discrete}.  
    The shear within the masked regions do not contribute to the peaks, so the
    height of the peaks will be underestimated.
  \item[Weighted:] Here we implement a weighting scheme which
    re-normalizes the filter $Q_{\rm NFW}$ to reflect the
    reduced contribution from masked pixels.  The integral in 
    equation \ref{Map_def} is replaced by the normalized sum:
    \begin{equation}
      M_{\rm ap}(\vec\theta_i) 
      = \frac{\sum_j 
        Q_{\rm NFW}(\vartheta_{ij})
        \gamma_t(\vec\theta_j)w(\vec\theta_j)}
      {\sum_j 
        Q_{\rm NFW}(\vartheta_{ij})w(\vec\theta_j)}
    \end{equation}
    where $w(\vec\theta_j) = 0$ if the pixel is masked, and 
    $1$ otherwise.  This should correct for the underestimation of peak
    heights seen in the unweighted case.
\end{description}
In order to facilitate comparison between this weighted definition of \Map 
and the normal definition used in the unmasked and unweighted cases, 
we normalize the latter by $\sum_j Q_{\rm NFW}(\vec\theta_{ij})$, 
which is a constant normalization across the field.

\begin{figure}
 \centering
 \plotone{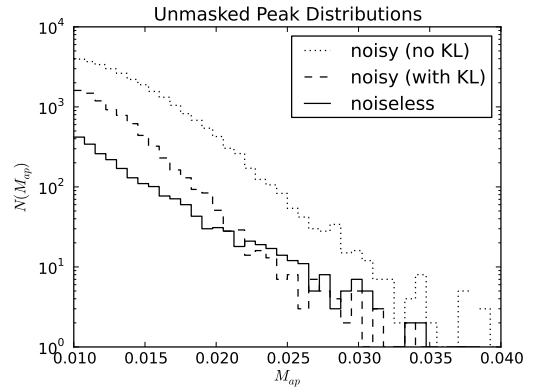}
 \caption{
   Comparison of the distribution of \Map peaks for unmasked shear,
   before and after filtering
   the field with KL (dotted line and dashed line, respectively).  The
   peak distribution in the absence of noise is shown for comparison
   (solid line).  It is clear that the addition of shape noise leads to 
   many spurious \Map peaks: noise peaks outnumber true peaks by nearly a
   factor of 10 for smaller peak heights.  Filtering by KL reduces these
   spurious peaks by about a factor of 3, and for larger peaks leads
   to a distribution similar in scale to that of the noiseless peaks.
   \label{fig_num_peaks}  
 } 
\end{figure}

Note that in both cases, it is the \textit{shear} that is masked, not the
\Map peaks.  Aperture mass is a non-local measure, so that the value can
be recovered even within the masked region.  This means that
masking will have a greater effect on the observed magnitude of the peaks
than it will have on the count.  In particular, on the small end of the
peak distribution, where the peaks are dominated by shape noise, the
masking of the shear signal is likely to have little effect on the
distribution of peak counts.  This can be seen in Figure~\ref{fig_num_peaks}.

\subsection{\Map Signal-to-Noise}
It is common in shear peak studies to study signal-to-noise peaks 
rather than directly study aperture-mass or convergence
peaks \citep[e.g.~][]{Wang09,Dietrich10,Schmidt10}.
We follow this precedent here.  
The aperture mass (Eqn.~\ref{Map_def_discrete}) is defined in terms of the
tangential shear.  Because we assume that the shear measurement is dominated
by isotropic, uncorrelated shape noise, the noise covariance of 
\Map can be expressed
\begin{eqnarray}
  [\Noise_M]_{ij} &\equiv& \langle M_{\rm ap}(\vec\theta_i) 
  M_{\rm ap}(\vec\theta_j)\rangle  \nonumber\\
  & = & \frac{1}{2}\sum_{k}
  Q^2_{\rm NFW}(\vartheta_{ik}) [\Noise_\gamma]_{kk} \delta_{ij}
\end{eqnarray}
where we have used the fact that shape noise is uncorrelated:
$[\Noise_\gamma]_{ij} = \langle n_i n_j\rangle \propto \delta_{ij}$.

In the case of a KL-reconstruction of a masked shear field, the reconstructed
shear has non-negligible correlation of noise between pixels.  
From Equations~\ref{shear_recons}-\ref{a_WF}, it can be shown that
\begin{eqnarray}
  \Noise_{\hat\gamma} 
  & \equiv  &
  \langle\myvec{\hat\gamma} \myvec{\hat\gamma}^\dagger\rangle\nonumber\\
  & = & \Noise_\gamma^{1/2}\myvec\Psi \mymat{M}^{-1} \mymat{\Psi}^\dagger \mymat{W}^2 \mymat{\Psi} \mymat{M}^{-1} \mymat{\Psi}^\dagger \Noise_\gamma^{1/2}
\end{eqnarray}
The covariance matrix $\Noise_{\hat\gamma}$ is no longer diagonal, but
the noise remains isotropic under the linear transformation, so that
$\Noise_{\hat{\gamma}_t} = \Noise_{\hat{\gamma}}/2$.
The aperture mass noise covariance can thus be calculated in a similar 
way to the non-KL case:
\begin{equation}
  [\Noise_M]_{ij} = \frac{1}{2}\sum_k\sum_\ell Q_{\rm NFW}(\vartheta_{ik})
  Q_{\rm NFW}(\vartheta_{j\ell}) [\Noise_{\hat\gamma}]_{k\ell}.
\end{equation}
This expression can be computed through standard linear algebraic techniques.  
The aperture mass signal-to-noise in each pixel is given by 
\begin{equation}
  [S/N]_i = M_{\rm ap}(\vec\theta_i)/\sqrt{[\mymat{\Noise_M}]_{ii}}
\end{equation}

\begin{figure}
 \centering
 \plotone{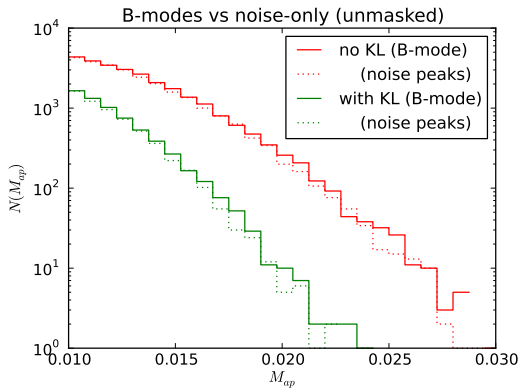}
 \caption{The comparison between B-mode peak distributions and the
   peak distributions for a shear field composed entirely of noise.
   As expected, the B-mode peak distributions are largely consistent
   with being due to noise only.  Because of this, we can use B-mode
   peaks as a rough proxy for the noise.
   \label{fig_B_noise} 
 }
\end{figure}

\begin{figure} 
 \centering
 \plotone{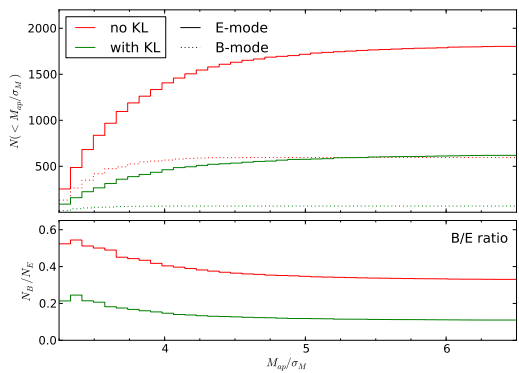}
 \caption{
   \textit{top panel:} The cumulative distributions of peaks in 
   signal-to-noise, for peaks with $M_{\rm ap}/\sigma_M > 3.25$.
   This is the statistic used by \citet{Dietrich10} to discriminate
   between cosmological models.  \textit{bottom panel:}  The ratio
   of B-mode to E-mode peak distributions.  Filtration by KL reduces
   the relative number of B-mode peaks by about 1/3.  Because B-mode
   peaks are a proxy for contamination by shape noise
   (see Figure~\ref{fig_B_noise}), this indicates that KL-filtration
   results in peak distributions less affected by statistical errors.
   KL also reduces the total number of both E and B peaks by about 2/3; 
   this effect can also be seen in Figure~\ref{fig_num_peaks}.
   \label{fig_EB_comp} 
 }
\end{figure}

\section{Discussion}
\label{Discussion}
\subsection{\Map Peak Distributions}
In Figures~\ref{fig_mask_nomask}-\ref{fig_EB_comp}
we compare the peak distribution obtained with and without KL.
We make three broad qualitative observations which point to the
efficacy of KL in interpolation of masked shear fields, and in
the filtration of shape-noise from these fields.  We stress the
qualitative nature of these results: quantifying 
these observations in a statistically rigorous way would require
shear fields from an ensemble of cosmology simulations, which is beyond the
scope of this work.  These results nevertheless point to the
efficacy of KL analysis in this context.

\textbf{KL filtration corrects for the bias due to masking.}
Figure~\ref{fig_mask_nomask} compares the effect of masking on the
resulting peak distributions with and without KL.  The left panel
shows the unmasked noisy peak distribution, and the masked
peak distributions resulting from the weighted and unweighted
approaches described in Section~\ref{MaskingEffects}.  
Neither method of accounting for the masking accurately 
recovers the unmasked distribution of peak heights. The unweighted
approach (green line) leads to an underestimation of peak heights.  
This is to be expected, because it does not correct for the missing 
information in the masked pixels.  The weighted approach, on the other
hand, over-estimates the counts of the peaks.  We suspect
this is due to an analog of Eddington bias: the lower signal-to-noise 
ratio of the weighted peak statistic leads to a larger scatter in peak heights.
Because of the steep slope of the peak distribution, this scatter 
preferentially increases the counts of larger peaks.  This suspicion is
confirmed by artificially increasing the noise in the unmasked peak function.
Increasing $\sigma_\epsilon$ from 0.30 to 0.35 in the unmasked case
results in a nearly identical peak function to the weighted masked case.

The right panel of Figure~\ref{fig_mask_nomask} shows that when 
KL is applied to the shear field, the distribution of the masked and
unmasked peaks is very close, both for E-mode and B-mode peaks.
This indicates the success of the KL-based interpolation outlined in
Section~\ref{KL_Interpolation}.  Even with 20\% of the pixels masked,
the procedure can recover a nearly identical peak distribution as 
from unmasked shear.

\textbf{KL filtration reduces the number of noise peaks.}
Comparison of the unmasked lines in the left and right panels of 
Figure~\ref{fig_mask_nomask} shows that application of KL to a shear
field results in fewer peaks at all heights.  This is to be expected:
when a reconstruction is performed with fewer than the total number
of KL modes, information of high spatial frequency is lost.  In this
way, KL acts as a sort of low-pass filter tuned to the particular
signal-to-noise characteristics of the data.  Figure~\ref{fig_num_peaks}
over-plots the KL and non-KL peak distributions with the noiseless
peak distribution.  From this figure we see that the inclusion of
shape noise results in nearly an order-of-magnitude more peaks than 
the noiseless case.  The effect of noise on peak counts lessens slightly 
for higher-\Map peaks: this supports the decision of \citet{Dietrich10} 
to limit their distributions to peaks with a signal-to-noise 
ratio greater than 3.25: the vast majority of peaks are lower 
magnitude, and are overwhelmed by the effect of shape noise.

Omission of higher-order KL modes of shear field reduces the number of
these spurious peaks by a factor of 3 or more.  For low-magnitude peaks,
$(M_{\rm ap}\lesssim 0.02)$, KL still produces peak counts which are 
dominated by noise. For higher-magnitude peaks, the number of observed 
KL peaks more closely approaches the number of peaks in the noiseless case.

\textbf{KL filtration reduces the presence of B-modes.}
To first order, weak lensing shear is expected to consist primarily of
curl-free, E-mode signal.  Because of this, the presence of B-modes can
indicate a systematic effect.  It is not obvious that filtration by KL
will maintain this property: as noted in Section~\ref{WhichCorrelation},
KL modes individually are agnostic to E-mode and B-mode information.  
E\&B information is only recovered within a 
complex-valued linear combination of the set KL modes.

Figure~\ref{fig_B_noise} shows a comparison between the unmasked
B-mode peak functions from Figure~\ref{fig_mask_nomask} and the associated
E-mode peak functions due to shape-noise only.  For both the non-KL version
and the KL version, the B-mode peak distributions closely follow the
distributions of noise peaks.  This supports the use of B-mode peaks as
a proxy for the peaks due to shape noise, even when truncating higher-order
KL modes.

The near-equivalence of B-modes and noise-only peaks shown in 
Figure~\ref{fig_B_noise} suggests a way of recovering the true peak function,
by subtracting the B-mode count from the E-mode count as a proxy for the 
shape noise. This approach has one fatal flaw: because it involves computing
the small difference between two large quantities, the result has
extremely large uncertainties.  It should be
noted that this noise contamination of small peaks is not an impediment to
using this method for cosmological analyses: the primary information
in shear peak statistics is due to the high signal-to-noise peaks.

In the top panel of Figure~\ref{fig_EB_comp}, we show the cumulative 
distribution of peaks in signal-to-noise, for peaks with $S/N > 3.25$:
the quantity used as a cosmological discriminant in \citet{Dietrich10}.
The difference in the total number of E-mode peaks in the KL and non-KL
approaches echoes the result seen
in Figure~\ref{fig_num_peaks}: truncation of higher-order KL modes acts as
a low-pass filter, reducing the total number of peaks by a factor of $\sim 3$.
More interesting is the result shown in the lower panel of 
Figure~\ref{fig_EB_comp}, where the ratio of B-mode peak counts to E-mode 
peak counts is shown.  
Before application of KL, the B-mode contamination is above 
30\%.  Filtration by KL reduces this contamination by a factor of $\sim 3$,
to about 10\%.  This indicates that the truncation of higher-order KL modes
leads to a preferential reduction of the B-mode signal, which traces
the noise.  This is a promising observation: the counts of
high signal-to-noise peaks, 
which offer the most sensitivity to cosmological parameters 
\citep{Dietrich10}, are significantly less contaminated by noise after 
filtering and reconstruction with KL.  This is a strong indication that 
the use of KL could improve the cosmological 
constraints derived from studies of shear peak statistics.

Note that in Figure~\ref{fig_EB_comp} we omit the masked 
results for clarity.  The masked cumulative signal-to-noise peak 
functions have B/E ratios comparable to the unmasked versions, so 
the conclusions here hold in both the masked and unmasked cases.

\subsection{Remaining Questions}
The above discussion suggests that KL analysis of masked shear fields holds
promise in constraining cosmological parameters of shear peaks in both
masked and unmasked fields.  KL greatly
reduces the number of spurious noise peaks at all signal-to-noise levels.
It minimizes the bias between masked and unmasked constructions, and leads
to a factor of 3 suppression of the B-mode signal, which is a proxy
for the spurious signal introduced through shape noise.

The question remains, however, how much cosmological information is contained
in the KL peak functions.  The reduction in level of noise peaks is promising,
but the omission of higher-order modes in the KL reconstruction leads to 
a smoothing of the shear field on scales smaller than the cutoff mode.
This smoothing could lead to the loss of cosmologically useful information.
In this way, the choice of KL mode cutoff can be thought of as a balance 
between statistical and systematic error.
The effect of these competing properties on cosmological
parameter determination is difficult to estimate.  Quantifying this effect
will require analysis within a suite of synthetic shear maps, similar to the
approach taken in previous studies \citep[e.g.][]{Dietrich10,Kratochvil10}, 
and will be the subject of future work.

Another possible application of KL in weak lensing is to use KL to 
directly constrain 2-point information in the measured shear data.  
In contrast to the method outlined in the current work, 
KL basis functions can be computed for the unmasked region only.
The projection of observed data onto this basis can be used to 
directly compute cosmological parameters via the 2-point function, 
without ever explicitly calculating the power spectrum.  
This is similar to the approach taken for galaxy counts
in \citet{Vogeley96}.  We explore this approach in a followup work.
 
{\it Acknowledgments:} 
We are very grateful to Risa Wechsler, Michael Busha, and Matt Becker for 
providing us simulated shear catalogs.  The simulation used to produce the 
catalogs is one of the Carmen simulations, a 1 Gpc simulation run by M. Busha 
as part of the LasDamas project. 
We thank Alex Szalay for helpful discussions regarding KL analysis in 
cosmological contexts.
Support for this research was provided by DOE Grant DESC0002607.
This research was partially supported by NSF AST-0709394 and NASA NNX07-AH07G.

\bibliography{shear_KL}

\begin{appendix}
\section{Choice of KL Parameters}
\label{Choosing_Params}
The KL analysis outlined in Section~\ref{KL_Intro}
has only two free parameters: the number of
modes $n$ and the Wiener filtering level $\alpha$.  Each of these parameters
involves a trade-off: using more modes increases the amount of information
used in the reconstruction, but at the expense of a decreased signal-to-noise
ratio.  Decreasing the value of $\alpha$ to $0$ reduces the smoothing effect 
of the prior, but can lead to a nearly singular convolution matrix 
$\mymat{M}_{(n,\alpha)}$, 
which results in unrealistically large shear values in the poorly-constrained 
areas areas of the map (i.e.~masked regions).

To inform our choice of the number of modes $n$, we recall the trend of 
spatial scale with mode number seen in Figure~\ref{fig_bandpower}.  Our 
purpose in using KL is to allow interpolation in masked regions.  To this 
end, the angular scale of the mask should inform the choice of angular
scale of the largest mode used.  An eigenmode which probes scales much smaller
than the size of a masked region will not contribute meaningful information to
the reconstruction within that masked region.  Considering the pixels within
our mask, we find that 99.5\% of masked pixels are within 2 pixels of a
shear measurement.  This corresponds to an angular scale of $\ell=6140$.
Consulting Figure~\ref{fig_bandpower}, we see that modes larger than
about $n=900$ out of 4096 will probe length scales significantly 
smaller than the mask scale.  
Thus, we choose $n=900$ as an appropriate cutoff for our reconstructions.

To inform our choice of the Wiener filtering level $\alpha$, we examine the
agreement between histograms of \Map peaks for a noise-only DES field
with and without masking (see Section~\ref{Shear_Peaks}).  
We find that for large (small) values of $\alpha$, the number
of high-\Map peaks is underestimated (overestimated) in the masked 
case as compared to the unmasked case.  
Empirically, we find that the two agree at $\alpha = 0.15$; 
we choose this value for our analysis.  Note that this 
tuning is done on noise-only reconstructions, 
which can be generated for observed data by assuming that
shape noise dominates: 
\begin{equation}
  [\Noise_\gamma]_{ij} = \frac{\sigma_\epsilon}{n_i^2}\delta_{ij}.
\end{equation}
The $\alpha$-tuning can thus be performed on artificial noise realizations 
which match the observed survey characteristics.

We make no claim that $(n,\alpha) = (900,0.15)$ is the optimal
choice of free parameters for KL: determining this would involve a more
in-depth analysis.  They are simply well-motivated choices which we use to
make a case for further study.
\end{appendix}

\end{document}